\documentclass[aps, prb, reprint, floatfix, superscriptaddress]{revtex4-1}
    \usepackage{amsmath}
    \usepackage{amssymb}
    \usepackage{graphics}
    \usepackage{graphicx}
    \usepackage{color}
    \usepackage{physics}    
    \usepackage[colorlinks=true, citecolor=blue, urlcolor=blue, linkcolor=black]{hyperref}
    \usepackage[capitalize]{cleveref}
    \usepackage[detect-weight=true, detect-family=true, separate-uncertainty=true, multi-part-units=single]{siunitx}

\begin{document}
\title{Observation of roton emission from a quantized vortex}

\author{A.\,Lester}
    \affiliation{Department of Physics, Lancaster University, Lancaster, LA1 4YB, United Kingdom}
\author{N.\,Morrison}
    \affiliation{Department of Physics, Lancaster University, Lancaster, LA1 4YB, United Kingdom}
    \affiliation{Now at: Department of Physics, University of California, Berkeley, USA}
\author{F.\,Novotny}
\author{D.~Schmoranzer}
    \affiliation{Department of Low Temperature Physics, Charles University, 121 16 Praha 2, Czech Republic}
\author{S.\'{O}\,Peat\'{a}in}
    \affiliation{Department of Physics, Lancaster University, Lancaster, LA1 4YB, United Kingdom}
    \affiliation{Now at: Centre for Quantum Technologies, National University of Singapore, Singapore.}
\author{V.\, Zavjalov}
    \affiliation{Department of Physics, Lancaster University, Lancaster, LA1 4YB, United Kingdom}
    \affiliation{Now at: IQM Quantum Computers, Keilaranta 19 D, 02150 Espoo, Finland}
\author{V.~Tsepelin}
    \affiliation{Department of Physics, Lancaster University, Lancaster, LA1 4YB, United Kingdom}
\author{S.~Kafanov}\email{sergey.kafanov@gmail.com}
    \affiliation{Department of Physics, Lancaster University, Lancaster, LA1 4YB, United Kingdom}

\maketitle






\textbf{
    Turbulence in inviscid quantum fluids offers unparalleled access to the universal principles of non-equilibrium dynamics, spanning a vast range of length scales from macroscopic flow down to the individual vortex core. In the zero-temperature limit, the microscopic mechanism by which the turbulent energy cascade terminates in the absence of viscosity remains a foundational challenge in quantum hydrodynamics. While prevailing theoretical descriptions prioritize phonon emission, they fail to account for the strong interatomic correlations that give rise to the roton minimum in superfluid $^4\mathrm{He}$. Here, we report the direct observation of roton emission from a single quantized vortex using a high-quality-factor nanomechanical resonator at 10 mK. We identify a sharp onset of dissipation at a critical velocity, and measure the energy loss per cycle, which corresponds quantitatively to the roton gap energy. Our findings address the long-standing mystery of zero-temperature energy relaxation by establishing roton emission as the primary dissipation channel in strongly correlated quantum liquids.}


Quantum turbulence, the intricate flow of superfluids, provides a unique platform for investigating the universal principles of non-equilibrium fluid dynamics~\cite{barenghi_quantum_2023,tsubota_quantum_2025}. In the zero-temperature limit, where viscous dissipation vanishes, energy cascades from macroscopic scales down to individual quantized vortices through a Richardson-like mechanism. Although the large-scale statistics of this flow mirror classical Kolmogorov turbulence~\cite{PhysD.237.2195, PhysRevLett.89.145301, NaturePhys.7.2011}, the microscopic mechanism by which the cascade terminates remains a key challenge in quantum hydrodynamics. Over the last decade, high-temperature visualization experiments in superfluid \(^4\mathrm{He}\) have captured reconnections between individual vortices\cite{PNAS.105.13707}, the propagation of vortex rings\cite{Tang2023} and the excitation of Kelvin waves~\cite{ProcNatAcadSci.111.4707}, which are helical excitations moving along vortex cores. These observations have firmly established that energy at the intervortex scale is redistributed via such topological events. 

Recent breakthroughs have enabled the systematic study of Kelvin waves in both classical water~\cite{BarckickeNature2026} and superfluid helium~\cite{NatPhys.21.233}, providing definitive confirmation of the expected dispersion relation. However, the subsequent Kelvin wave cascade, nonlinear interactions of Kelvin waves, and the ultimate decay mechanism of individual vortices into the bulk fluid have remained experimentally elusive. In previous superfluid experiments, the interaction of the vortex core with the bulk excitations (the normal fluid) dominated the decay, thereby masking the fundamental dissipation events critical for establishing a self-consistent theory of quantum turbulence. Consequently, this final stage of energy relaxation has been accessible only through theoretical models and numerical simulations. 

Standard theoretical descriptions, particularly those based on the Gross-Pitaevskii equation (GPE), attribute the ultimate dissipation to the Kelvin wave cascade terminating via the emission of elementary excitations, predominantly phonons (sound waves) at wavenumbers comparable to the inverse core size~\cite{PhysRevB.64.134520, PhysRevLett.92.035301}. However, this framework remains incomplete for superfluid \(^4\mathrm{He}\). The GPE models a weakly interacting Bose gas and ignores the strong interactions that give rise to the roton minimum in the helium excitation spectrum. This minimum governs the breakdown of superfluidity at the Landau critical velocity. For an object travelling in superfluid helium, phonon emission is expected only when the flow velocity exceeds the speed of sound~\cite{tsubota_quantum_2025}, which is approximately three times higher than the Landau threshold. Consequently, rotons should lead the dissipation~\cite{JLTP.128.167}. This reasoning is supported by a recent GPE model incorporating the roton minimum~\cite{PhysRevB.105.014515}, in which the simulations demonstrate that roton emission is the predominant dissipation channel above a critical flow velocity, accompanied by the generation of vortices. Furthermore, advanced microscopic theories and generalized non-local models suggest that the vortex core in $^4$He is not a simple density depletion, but a structured ``cloud'' of virtual bound states, primarily rotons. Under dynamic perturbations, these virtual excitations are predicted to destabilize, converting into real, propagating rotons~\cite{PhysRevLett.121.015302}. The latter implies that quantized vortices act as non-thermal sources of elementary excitations, a phenomenon conceptually akin to vacuum decay in high-energy physics.

Here, we report an experimental probing of single-vortex dynamics in superfluid $^4$He and demonstrate the emission of rotons from a quantum vortex. Using a high-quality-factor nanomechanical resonator, we trap and drive a single quantum vortex to probe its relaxation mechanism. We identify a distinct breakdown of dissipationless motion at a critical velocity, where the energy loss per cycle corresponds quantitatively to the emission of one or two rotons at higher excitation drives. Our findings provide direct experimental evidence that challenges the phonon-dominated view of vortex-line relaxation and establishes roton emission as a primary dissipation mechanism for a single driven vortex in superfluid $^4$He. These results support the theoretical prediction that quantized vortices can serve as non-thermal sources of roton emission.

\begin{figure}
\includegraphics[width=\linewidth]{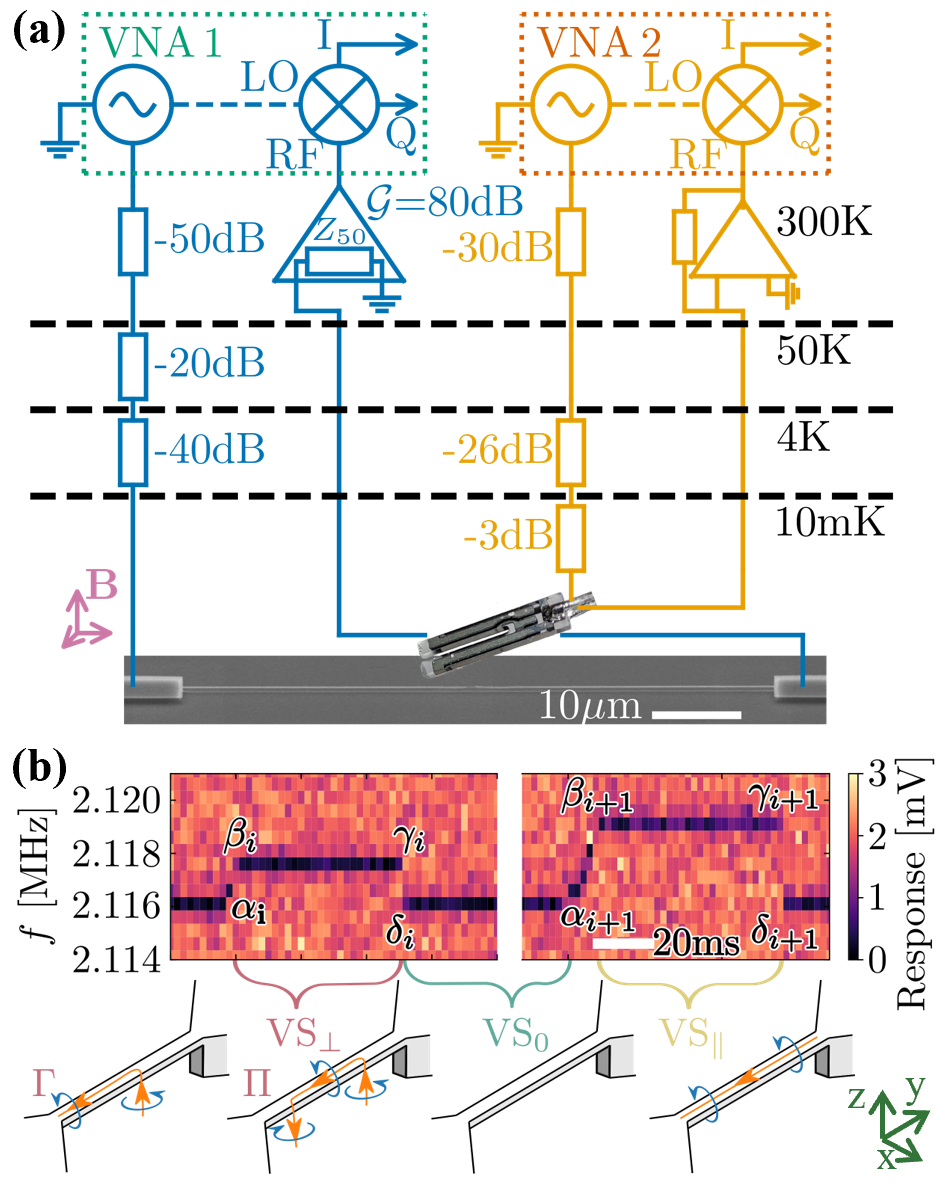}
\caption{\textbf{(a)} The nanobeam’s actuation and detection circuit (blue) is driven by a vector network analyzer (VNA) via a series of attenuators installed at various cryostat temperature stages. The transmission signal is amplified at room temperature. The quartz tuning fork, positioned \(\sim\)\SI{2}{\milli\meter} above the nanobeam, generates quantum turbulence. The tuning fork is driven and read out using a second VNA circuit (orange). \newline \textbf{(b)} The colourmap shows time evolution of two examples of vortex capture and annihilation events recorded using a multi-frequency lock-in amplifier~\cite{JLowTempPhys.184.1080}. Darker regions indicate the location of the beam's resonance. The lowest resonance value near \(\sim\)\SI{2.116}{\mega\hertz} corresponds to the vortex-free state (\(\text{VS}_0\)) observed in the absence of vortices~\cite{NatComm.12.2645}. The trapped vortex alters the beam’s tension and increases its resonance frequency. The left panel, with the highest frequency shift (approximately \SI{3}{\kilo\hertz}) depicts the most commonly occurring event of a fully trapped vortex, aligned along the entire length of the nanobeam (\(\text{VS}_\parallel\))~\cite{NatComm.12.2645}. Right panel illustrates capture and subsequent annihilation of a vortex in the perpendicular configuration (\(\text{VS}_{\perp}\)). Cartoons below the colourmap sketch the vortex configurations, including a single (\(\text{VS}_{\perp_\Gamma}\)) and double (\(\text{VS}_{\perp_\Pi}\)) vortex branches connecting the beam to the substrate. \label{fig:Circuit}}
\end{figure}

To resolve the microscopic energy dissipation of a single quantum vortex, we utilised a doubly-clamped nanobeam resonator (\(\ell = \SI{70}{\micro\meter}\), \(\qtyproduct{200x130}{\nano\meter}\) cross-section) suspended \(\sim\SI{2}{\micro\meter}\) above a silicon substrate (\cref{fig:Circuit}a). The nanobeam was driven and measured magnetomotively using a vector network analyzer (VNA). Homodyne detection provided both amplitude and phase information of the device motion. In vacuum at 10 mK, the nanobeam exhibits a high internal quality factor (\(Q_\text{mech} = \num{5e5}\)) and linear behaviour at low amplitudes, providing a baseline for detecting dissipation events in superfluid \(\mathrm{{}^{4}He}\).

We generated a vortex tangle in the surrounding superfluid $^4$He using a quartz tuning fork~\cite{NatComm.12.2645}. By monitoring the nanobeam's resonance frequency in real-time (\cref{fig:Circuit}b), we observed discrete frequency shifts, signalling the trapping of individual vortex loops~\cite{NatComm.12.2645}. We categorize these events into three distinct configurations: the vortex-free state (\(\text{VS}_0\)); the fully trapped parallel state (\(\text{VS}_\parallel\)), where the vortex aligns along the entire length of the beam; and the perpendicular state (\(\text{VS}_\perp\)), where the vortex segment aligns along a part of the beam and connects it to the substrate.

\begin{figure*}
\includegraphics[width=\linewidth]{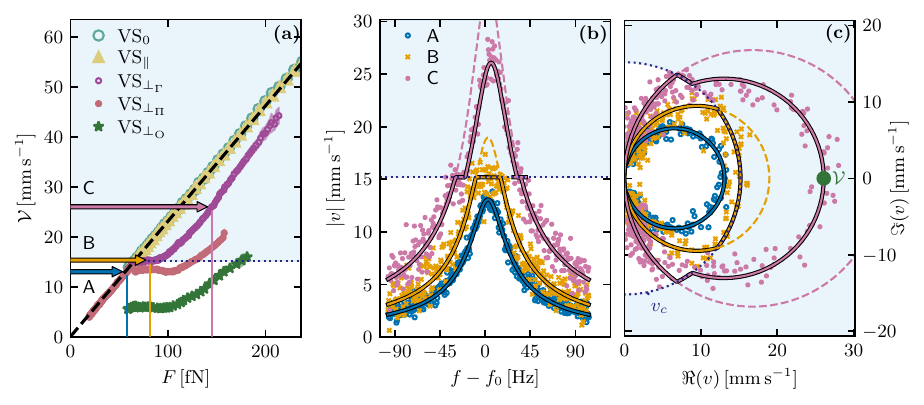}
\caption{\textbf{(a)} Nanobeam resonant velocity amplitude, \(\mathcal{V}\), as a function of driving force amplitude. Three distinct dynamical regimes, (A), (B), and (C) for the \(\text{VS}_{\perp_\Gamma}\) state, are colour-coded and labelled across all subplots and \cref{fig:Energy}. Region~(A) corresponds to the low-damping regime, where the quality factor (\(Q = Q_\text{mech}\)) is limited by beam's intrinsic mechanical losses. In region~(B), the peak velocity saturates at the critical value \(v_c\), indicating the onset of additional dissipation. Region~(C) denotes the higher-damping regime, where the velocity exceeds \(v_c\), and the system shows additional energy loss compared to region (A). Arrows on the (\(\text{VS}_{\perp_\Gamma}\)) curve mark the actuation amplitudes corresponding to the Nyquist plots and velocity spectra shown in panels~(b) and~(c), respectively. \textbf{(b)} Frequency spectra of the nanobeam velocity, centred at the resonance frequency \(f_0 = \SI{2.1876447}{\mega\hertz}\). \textbf{(c)} Nyquist diagrams of the nanobeam velocity. Solid and dashed circles represent fits for the lowest- and higher-damping regimes, respectively. The dotted semicircle indicates the critical velocity \(v_c = \SI{15.20 \pm 0.06}{\milli\meter\per\second}\). The resonant velocity amplitude, \(\mathcal{V}\), is given by the intersection of the outer fitted circle with the real axis. Measurements were taken at around \SI{10}{\milli\kelvin}.\label{fig:FV} }
\end{figure*}

A crucial control experiment is the response of the fully trapped state ($\text{VS}_{\parallel}$). As shown in (\cref{fig:FV}(a)), both the vortex-free (\(\text{VS}_0\)) and parallel (\(\text{VS}_\parallel\)) configurations exhibit a linear response with a quality factor indistinguishable from the vacuum value across the entire driving range. This result is significant: it demonstrates that the mere presence of a superfluid and a vortex does not introduce drag and, more importantly, that acoustic dissipation (phonon emission) is negligible at these oscillation frequencies. 

In stark contrast to the parallel configuration, the perpendicular states (\(\text{VS}_\perp\)) reveal a highly nonlinear dissipation profile governed by a critical velocity threshold. We identify three distinct dynamical regimes (\cref{fig:FV}a): (A) a linear regime at low drives, where the response remains identical to the \(\text{VS}_0\) and \(\text{VS}_\parallel\) states; (B) a nonlinear regime with the onset of dissipation, where, as the drive increases, the velocity amplitude abruptly saturates at a critical value. The response curve deviates from a Lorentzian profile (\cref{fig:FV}b), and the Nyquist plot transitions from a circle to a flattened semi-circle (\cref{fig:FV}c), a hallmark of velocity-limited damping.; (C) a nonlinear regime above the critical velocity leading to increased damping with respect to regime A and eventual depinning and loss of the vortex.

The saturation in regime B indicates the activation of a specific dissipation channel that limits the velocity~\cite{PhysFluids.37.031305}. The activation threshold is sharp, ruling out continuous damping mechanisms such as viscous drag or acoustic radiation. While Kelvin wave propagation is continuous and dispersive, as recently visualized in 3D by Minowa et al.~\cite{NatPhys.21.233}, nonlinearities may give rise to the appearance of a Kelvin-wave cascade. The latter exhibits a strong energy dependence on the wave amplitude~\cite{JetpLett.111.389}, which should result in a constant velocity plateau but is unlikely to transition to regime C. 

\begin{figure}
\includegraphics[width=\linewidth]{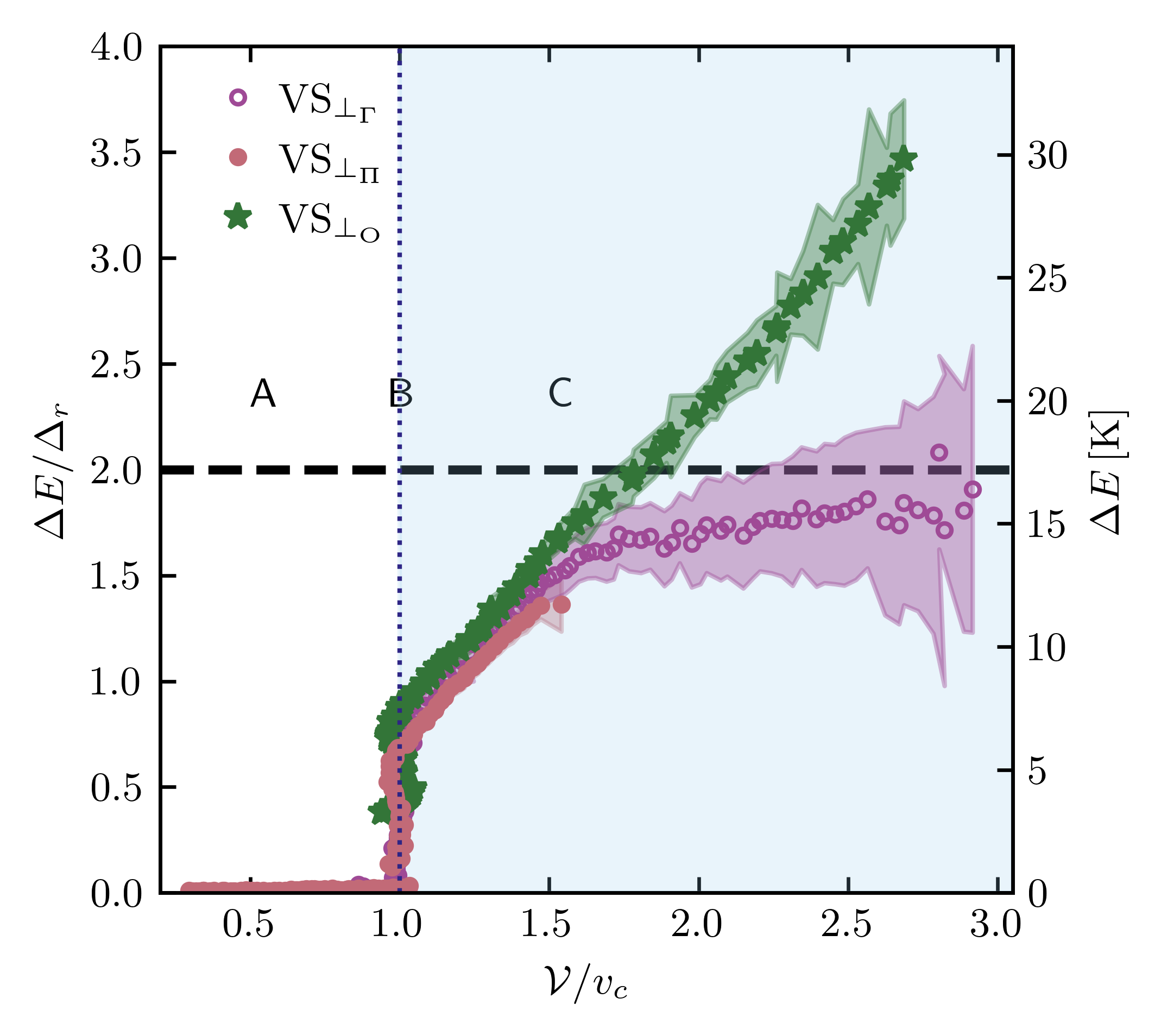}
\caption{Energy loss per oscillation cycle for typical \(\mathrm{VS}_{\perp}\) configurations. The \(\text{VS}_{\perp_\Gamma}\) event corresponds to a \(\Gamma\)-shaped configuration, \textit{i.e.}, a single vortex branch connects the nanobeam to the substrate. The \(\text{VS}_{\perp_\Pi}\) event corresponds to a \(\Pi\)-shaped configuration, \textit{i.e.}, where two vortex branches connect the nanobeam to the substrate. The energy values for the latter event have been scaled by a factor of 1/2. The \(\text{VS}_{\perp_\text{O}}\) event exhibits a continuous (non-quantized) energy dissipation mechanism, attributed to the continuously growing size of emitted vortex rings. The corresponding critical velocities for these configurations are: \(v_{\text{c},\,\Gamma} = \SI{15.20 \pm 0.06}{\milli\meter\per\second}\), \(v_{\text{c},\,\Pi} = \SI{13.53 \pm 0.06}{\milli\meter\per\second}\), \(v_\text{c,O} = \SI{5.98 \pm 0.03}{\milli\meter\per\second}\). \label{fig:Energy}}
\end{figure}

To identify the physical origin of the energy threshold, we analyzed the power dissipated by the vortex in regime B. \Cref{fig:Energy} displays the energy loss per oscillation cycle for various \(\text{VS}_{\perp}\) events. We observed that for the single-branch configuration ($\text{VS}_{\perp\Gamma}$, \(\text{VS}_{\perp_\text{O}}\)), the energy loss plateaus at a value quantitatively consistent with the energy required to emit a roton ($\Delta$, where $\Delta$ is the roton gap). Furthermore, we detected events ($\text{VS}_{\perp\Pi}$) where two vortex branches connected to the resonator; these showed double the dissipation of the single-branch events (\cref{fig:Energy}), supporting the roton emission process.

Regime C exhibits energy dissipation beyond the critical velocity plateau at higher beam drives. Measurements of the ($\text{VS}_{\perp\Gamma}$) and ($\text{VS}_{\perp\Pi}$) events reveal a gradual reduction in the damping force and a tendency to revert towards the behaviour in regime A. Such asymptotic behaviour is expected when a constant, fixed loss of energy becomes negligible compared to the total energy of the oscillator. The observed reduction is consistent with the quantized energy emission of two rotons per period per vortex branch (\cref{fig:Energy}). Ideally, one might expect the emission of two rotons per period, one during each half-cycle, immediately past the critical velocity plateau of regime B. However, in our case, the beam has trapped part of the vortex along its length, which results in an asymmetry induced by the Magnus force; consequently, one half-cycle reaches a lower peak velocity that initially remains below the threshold for excitation emission. While the periodic motion of the beam can excite Kelvin waves propagating between the beam and substrate, the energy loss via acoustic emission due to Kelvin waves in the absence of a cascade is several orders of magnitude below the damping we measure\cite{PhysRevB.64.134520}. The development of a Kelvin-wave cascade driven by nonlinear interactions would instead produce significant damping with a high-amplitude power-law dependence,\cite{JetpLett.111.389} forcing the system to remain in regime B. Furthermore, the emission of small vortex rings should result in constant damping rather than constant energy, as the ring diameter scales with the amplitude of the vortex motion. We observe constant damping behaviour in the (\(\text{VS}_{\perp_\text{O}}\)) state, which is consistent with non-quantized emission, such as the production of small vortex rings.   


The observed onset and quantization of energy losses from a single vortex line, trapped between the beam and the substrate, provide experimental evidence that roton emission is a primary mechanism for vortex decay. This finding aligns seamlessly with the established excitation framework of superfluid $^4\mathrm{He}$ and supports recent modified GPE simulations incorporating the roton minimum~\cite{PhysRevB.105.014515}. Furthermore, the emission of quantized energies, quantitatively matching those of a roton, corroborates theoretical predictions regarding the internal structure of the vortex core in superfluid $^4$He. Under dynamic perturbations, the virtual states formed due to the highly structured density modulations of the core are expected to undergo conversion into real, propagating rotons~\cite{PhysRevLett.77.5401, PhysRevLett.121.015302}. Overall, our observations address the long-standing puzzle of dissipation in the zero-temperature limit and bring us closer to a self-consistent picture of quantum turbulence.


All data used in this paper are available at http://dx.doi.org/10.17635/lancaster/researchdata/xxx, including descriptions of the data sets. 


The research leading to these results was supported by UKRI EPSRC and STFC (Grants ST/T006773/1, EP/P022197/1 and EP/X004597/1), as well as the European Union’s Horizon 2020 Research and Innovation Programme under Grant Agreement No. 824109 (European Microkelvin Platform). D.S. acknowledges support from the Czech Science Foundation, Grant No. 24-12253S. We are greatly thankful to L.~Melnikovskiy, A.~Golov, L.~Skrbek, and V.~Eltsov for valuable discussions that contributed to a better understanding of the described phenomenon. We gratefully acknowledge M.\,T.~Noble for selecting the vortex trapping events from the dataset of~\cite{NatComm.12.2645} used in~\cref{fig:Circuit}(b); Prof.\,Yu.\,A.~Pashkin (Lancaster University) for providing access to the dilution refrigerator; and the MSU team, led by Dr.\,V.\,A.~Krupenin, for supplying the nanomechanical resonator samples initially used in this work.


The experiment was designed by S.K. The cryogenic setup was developed and fabricated by S.O.P. and S.K. Low-temperature measurements were carried out by A.L., N.M., F.N., and S.K. Data analysis was performed by V.T., A.L., N.M., F.N., V.Z., D.S., and S.K. Interpretation of the results was conducted by V.T., D.S., and S.K.. The manuscript was primarily written by A.L., V.T., D.S., and S.K.




\bibliography{bibliography}

\section*{Methods}

\subsection{Device Description}

The nano-electromechanical system (NEMS) device consists of a doubly clamped aluminum-on-silicon nitride (Al-on-\(\mathrm{Si_3N_4}\)) composite nanobeam. The beam dimensions are defined lithographically, with a length \(\ell = \SI{70}{\micro\meter}\) and a width \(w = \SI{200}{\nano\meter}\). The \SI{100}{\nano\meter}-thick \(\mathrm{Si_3N_4}\) layer determines the mechanical properties of the beam, while the aluminum layer enables magnetomotive actuation and detection of its motion. The combined thickness of the aluminum and silicon nitride layers is \(t = \SI{130}{\nano\meter}\), corresponding to an effective density of \SI{3062}{\kilo\gram\per\meter\cubed}. The vacuum resonance frequency of the fundamental flexural mode is measured to be \(f_0 = \SI{2.238}{\mega\hertz}\). The experiment is performed in a brass cell filled with superfluid $^4$He at a temperature of \SI{10}{\milli\kelvin}, mounted on the mixing chamber of a cryogen-free dilution refrigerator.

\subsection{Measurement Scheme}
			
We actuated and detected the nanobeam motion using a magnetomotive detection scheme based on the frequency-dependent transmission response measured by a vector network analyzer (VNA). The AC excitation signal, \(V_\text{out}\), from the VNA was attenuated at key temperature stages of the cryostat, with a total attenuation of \(\mathcal{A}=\SI{-110}{dB}\), before passing through the nanomechanical resonator. In the presence of a perpendicular magnetic field \(B = \SI{1}{\tesla}\), the resulting current generated a Lorentz force that drove the mechanical motion of the device. The transmitted signal, which included a Faraday voltage component induced by the nanobeam’s motion, was amplified by \(\mathcal{G}=\SI{77}{dB}\) using a room-temperature amplifier with an input impedance \(Z_{50}=\SI{50}{\ohm}\), before acquisition by the VNA. The chosen magnetic field strength ensured that magnetomotive damping remained small compared to the intrinsic mechanical damping of the device (\(Q^{-1}_\text{mech}\))~\cite{PhysRevB.100.020506}.

\begin{figure}
\includegraphics[width=\linewidth]{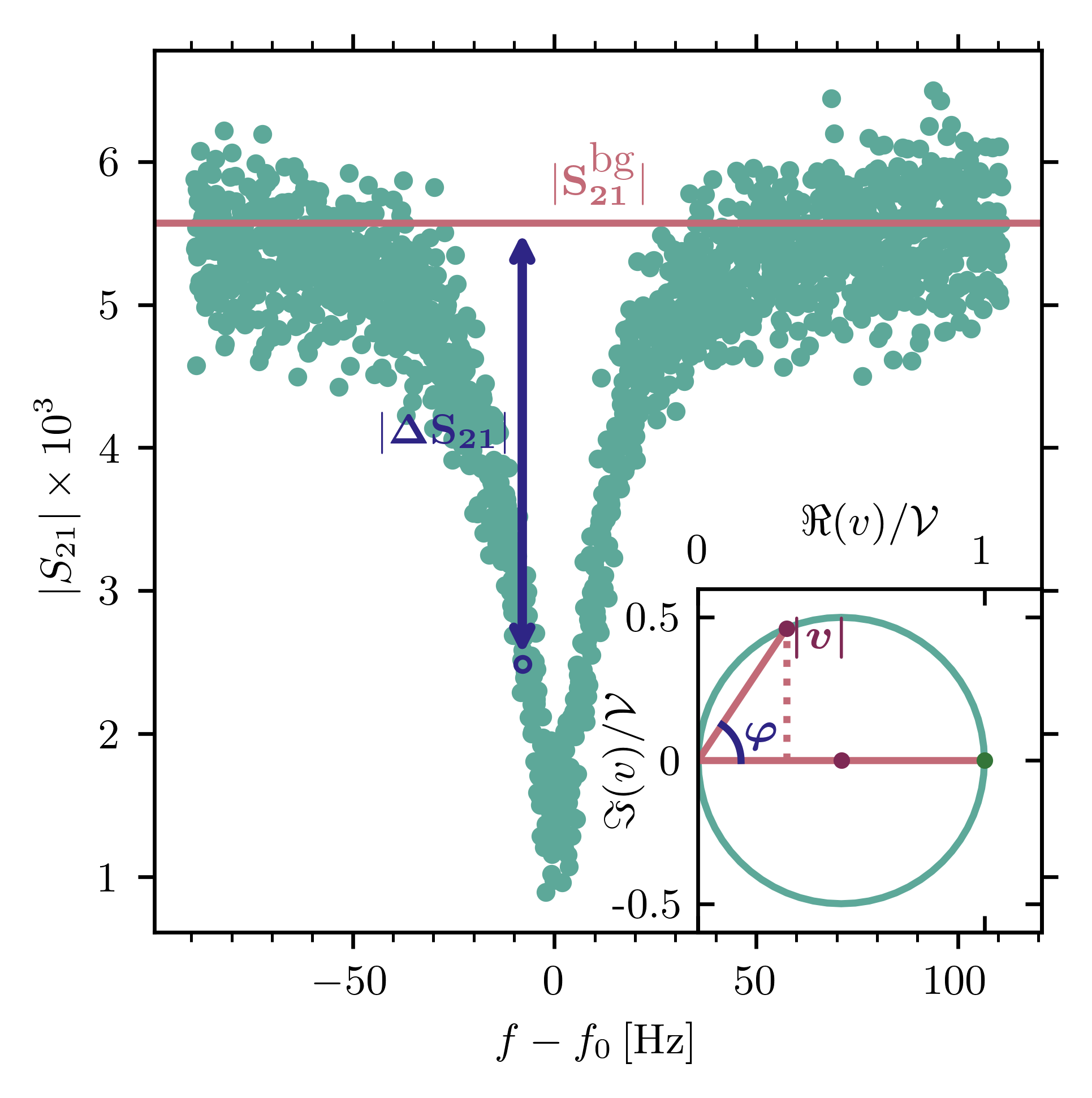}
\caption{Example of measured transmission magnitude \(|S_{21}|\) as a function of frequency, obtained using a vector network analyzer (VNA). The key parameters used to calculate the Lorentz force and beam velocity are indicated. (inset) The theoretical Nyquist diagram illustrates the derivation of the energy loss expressions in the system. \label{fig:Resonance}}
\end{figure}

The measured transmission signal \(S_{21}\), an example of which is shown in \cref{fig:Resonance}, enables the extraction of key parameters describing the nanobeam motion. 
            
The Lorentz force is given by
\[
F = \xi \dfrac{S^{\text{bg}}_{21}}{\mathcal{G}} \dfrac{V_\text{out}}{Z_{50}} B \ell,
\]
where \(S^{\text{bg}}_{21}\) is the background signal and \(\xi=0.523\) is the geometric factor for the fundamental mode of a doubly clamped beam. The factor was determined numerically, utilizing an approach similar to the calculations of the effective mass~\cite{AnnPhys.339.181}.

The nanobeam velocity, inferred from generated Faraday electromotive force, is 
\[
v = \dfrac{2}{\xi B \ell} \dfrac{\Delta S_{21}}{S^{\text{bg}}_{21}} \dfrac{V_\text{out}}{\mathcal{A}},
\]
where \(\Delta S_{21}\) is the change in transmission due to the beam motion. 

\subsection{Energy Losses Analysis}

The energy dissipation of the nanostring was analyzed using the equation of the linear harmonic oscillator:
\[
\ddot{x}+\delta\dot{x} + \omega_0^2 x = \dfrac{F}{m}\mathrm{e}^{\mathrm{i}\omega t},         
\]
where \(F/m\) is the amplitude of the driving force normalized by the oscillator's effective mass \(m\), $\delta$ is the damping, and $\omega_0$ is the resonance frequency.

The complex velocity response to the driving force acting at the normalized frequency \(\tilde{\omega} = \dfrac{\omega}{\omega_0}\) and the corresponding Nyquist diagram for the oscillator with a quality factor \(Q = \dfrac{\omega_0}{\delta}\) are given by:
\[
v = \dfrac{\mathrm{i} \tilde{\omega}}{{Q}\left(1 - \tilde{\omega}^2\right) + \mathrm{i} \tilde{\omega}}\mathcal{V},
\]
where \(\mathcal{V}=\dfrac{F}{m \omega_0} Q\) is the velocity's resonant amplitude. 
            
The relative phase of the velocity response to the applied force is given by 
\[
\varphi = \arctan\left(\dfrac{1}{\tilde{\omega}}-\tilde{\omega}\right) Q.
\]

Velocity's magnitude at the drive frequency \(\omega\) is given by
\[
|v| =\mathcal{V}\cos\varphi.
\]

The average power loss of the oscillator driven at the frequency \(\omega\) is given by
\[
\begin{split}
    W & = \dfrac{\omega}{2\pi}\int\limits^{\tfrac{\omega}{2\pi}}_0 F\cos(\omega t) \Re\{v\} \cos(\omega t)\, \mathrm{d}t = 
\dfrac{1}{2}F\mathcal{V}\cos^2\varphi = \\
& = \dfrac{1}{2}F\mathcal{V}\dfrac{1}{1+Q^2\left(\dfrac{1}{\tilde{\omega}}-\tilde{\omega}\right)^2}
\end{split}
\]
The power supplied to the oscillator to carry the oscillations' velocity, \(\mathcal{V}\), at the resonance frequency will be greater by 
\[
\Delta W = \dfrac{1}{2}\mathcal{V}\left(F_{\text{VS}_\perp} - F_{\text{VS}_0}\right).
\]
This expression was utilized for calculations of the energy losses per cycle from the data presented as an example in \cref{fig:FV}(a).

\end{document}